\documentclass[conference]{IEEEtran}
\usepackage{graphicx}
\usepackage{amssymb,amsmath,epsfig,citesort,url,fixltx2e}
\usepackage{algorithm,algorithmic}
\usepackage[caption=false,font=footnotesize]{subfig}
\psfull

\IEEEoverridecommandlockouts    

\usepackage{wrapfig}

\def\beq        {\begin{equation}}
\def\eeq        {\end{equation}}
\def\bea        {\begin{eqnarray}}
\def\eea       {\end{eqnarray}}

\makeatletter
\newlength \figwidth
\if@twocolumn
\else
\fi
\makeatother

\title{Aerial UAV-IoT Sensing for Ubiquitous Immersive Communication and Virtual Human Teleportation\vspace{-0.2cm}}
\author{\IEEEauthorblockN{Jacob Chakareski\thanks{To appear @ INFOCOM 2017.}}\IEEEauthorblockA{Electrical and Computer Engineering Department, The University of Alabama\vspace{-0.3cm}}
}

\begin{document}

\maketitle

\begin{abstract}
We consider UAV IoT aerial sensing that delivers multiple VR/AR immersive communication sessions to remote users. The UAV swarm is spatially distributed over a wide area of interest, and each UAV captures a viewpoint of the scene below it. The remote users are interested in visual immersive navigation of specific subareas/scenes of interest, reconstructed on their respective VR/AR devices from the captured data. The reconstruction quality of the immersive scene representations at the users will depend on the sampling/sensing rates associated with each UAV. There is a limit on the aggregate amount of data that the UAV swarm can sample and send towards the users, stemming from physical/transmission capacity constraints. Similarly, each VR/AR application has minimum reconstruction quality requirements for its own session. We propose an optimization framework that makes three contributions in this context. First, we select the optimal sampling rates to be used by each UAV, such that the system and application constraints are not exceed, while the priority weighted reconstruction quality across all VR/AR sessions is maximized. Then, we design an optimal scalable source-channel signal representation that instills into the captured data inherent rate adaptivity, unequal error protection, and minimum required redundancy. Finally, the UAV transmission efficiency is enhanced by the use of small-form-factor multi-beam directional antennas and optimal power/link scheduling across the scalable signal representation layers. Our experiments demonstrate competitive advantages over conventional methods for visual sensing. This is a first-of-its-kind study of an emerging application of prospectively broad societal impact.
\end{abstract}


\vspace{-0.3cm}

\section{Introduction}
\vspace{-0.2cm}
Cyber-physical/human systems (CPS/CHS) are set to play an increasingly prominent and important role in our lives and society,
advancing research and technology across different disciplines at the same time \cite{NSF_CPS_CHS,NIST_CPS,ProcIEEE:12,ACM_TCPS,IEEE_THMS}. Virtual/augmented reality (VR/AR) and UAV swarms are two emerging CPS/CHS technologies of prospectively broad societal impact. VR/AR suspends our disbelief of being there (at a remote location), akin to {\em virtual human teleportation} \cite{ApostolopoulosCCKTW:12}. Its ongoing explosion onto the market and into the mainstream\footnote{As evidenced by the flurry of related devices/services/platforms appearing on the market, released broadly by electronics/device manufacturers and Internet companies \cite{HTC_Vive,SonyPlayStationVR,RicohTheta,SamsungGear360,MicrosoftHoloLens,MagicLeap,Oculus,GoogleCardboard,Facebook360,JumpVR,FortuneProjectTango:16}, and stunning investment/acquisition deals/growth rates \cite{WiredMagazine:13,NYT_MagicLeap:16,CB_Insights:16,DigiCapital:16}. It is expected that AR/VR revenue will hit \$150B by 2020, disrupting the mobile market along the way \cite{TechCrunch:15}.} is a forerunner of the opportunities lying ahead. In brief, by equipping us with the capacity to travel virtually and apply super-human-like vision, VR/AR can help us achieve a broad range of technological and societal advances\footnote{Seeing from multiple 3D perspectives, around/through obstacles, and in ways not possible before will lead to knowledge discovery that will considerably enhance human/machine perception and decision making in diverse tasks \cite{ServettiM:13,Masala:13,VasudevanZKLBN:10,SchreerFAEKB:08,KimHSHOTSK:05,WeidlichSW:08,SourinSS:00,SnavelySGSS:10,WuLLY:15,SoteloDMSB:WuLLY:12,WongYLP:13,YangCS:15}. These gains will stimulate societal-scale applications that will enhance energy conservation, quality of life, and the global economy.
}. UAVs can have a similar transformative impact on remote sensing applications, by lowering their cost/extending their scope \cite{MIT_TECH_REVIEW:14c}.


We envision a future where UAV-deployed VR/AR immersive communication will {\em break existing barriers} in remote sensing, monitoring, localization, navigation, etc. The thereby achieved advances will benefit diverse applications spanning the environmental/weather sciences, public/national safety, disaster relief, and transportation. By dispensing with the need for our physical presence (transportation) there, they will simultaneously {\em make impact on climate change} \footnote{According to the U.S. Environmental Protection Agency, human transportation contributes to 1/3 of our carbon emission footprint \cite{EPA_GHGEM:16}. Climate and weather disasters attributed to climate change alone cost the American economy more than \$100B in 2012 \cite{WH:14}.}.

\begin{figure}[htb]
\centering
   \includegraphics[width=\figwidth]{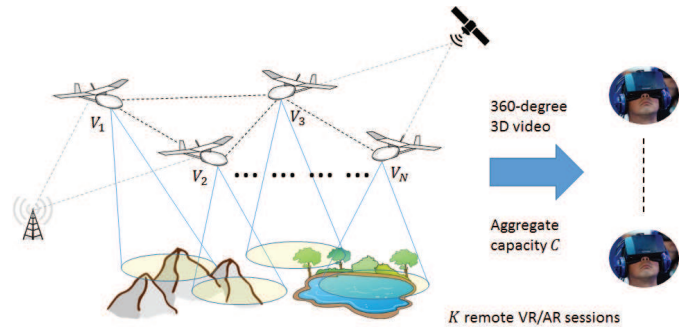}
   \caption{VR/AR immersive communication via UAV IoT aerial sensing.}
   \label{fig:ProblemSetup}
\end{figure}

In this paper, we study UAV IoT aerial sensing that delivers multiple VR/AR immersive communication sessions to remote users, as illustrated in Figure~\ref{fig:ProblemSetup}. The UAV swarm is spatially distributed over a wide area of interest, where each UAV $i$ records its own viewpoint $V_i$ of the scene below it. Each remote VR/AR user is interested in immersive navigation of a subarea of specific interest to him/her. The aggregate communication capacity of the sensing system is $C$ and needs to support the delivery of $K$ 360$^\circ$-navigable 3D video representations of the $K$ subareas of interest, constructed from the UAV-captured viewpoint data $\{V_1,\dots,V_N\}$.

The reconstruction quality of each subarea VR/AR visual representation will depend on the specific set of captured viewpoints and the sampling rate $R_i$ at which each $V_i$ has been acquired. Our objective will be to find the optimal set of sampling rates for the UAV swarm such that the aggregate priority-weighted reconstruction quality across all VR/AR sessions is maximized. At the same time, the minimum reconstruction quality requirements for the VR/AR sessions should be met, as well, with the selected optimal sampling rates. Furthermore, we design an optimal scalable source-channel signal representation that instills into the captured data inherent rate adaptivity, unequal error protection, and minimum required redundancy. Finally, the UAV transmission efficiency is enhanced by the use of small-form-factor multi-beam directional antennas and optimal power/link scheduling across the scalable signal representation layers.

Due to the complex interdependencies of these three system blocks, in this preliminary study, we analyzed and optimized each individually, while still keeping them coupled in their design and operation through the selected sampling/encoding rates that flow through each. Motivated by our preliminary results and advances, we plan to investigate a tighter integration of the three blocks into a unifying framework, as part of our ongoing/future work efforts, and analytically quantify the benefits versus drawbacks of such an approach. Similarly, we plan to derive further theoretical insights and analysis of our optimization methods as part of a follow-up study, which could not be included here due to limited space.

In the remainder, we first provide an overview of related work in Section~\ref{sec:RelatedWork}. Then, we describe our system modeling in Section~\ref{sec:SystemModel}, while in Section~\ref{sec:ProblemFormulation} we provide the problem formulation, subsequently. Analysis and optimization solution of the problem are provided in Section~\ref{sec:AnalysisOptimization}.
Our scalable source-channel signal representation is introduced in Section~\ref{sec:ScalableSourceChannel}, while our transmission power layered scheduling via multi-beam directional antennas is presented in Section~\ref{sec:LayeredScheduling}. Comprehensive experimental analysis and benchmarking are provided in Section~\ref{sec:Experiments}, and we conclude in Section~\ref{sec:Conclusion}.

\section{Related work}
\label{sec:RelatedWork}
VR/AR immersive communication via UAV IoT sensing is a new topic that has not been studied before. Closely related areas include ground-based multi-camera wireless sensing for multi-view systems \cite{Chakareski:15}, immersive telecollaboration \cite{SchreerFAEKB:08,VasudevanZKLBN:10,HosseiniK:15}, multi-view video coding/communication \cite{CheungOC:11,ChakareskiVS:12,Chakareski:14c,ChakareskiVS:15}, and 360$^\circ$ video streaming \cite{QianHJG:16,HosseiniS:16}. Another related area is graph-based signal processing for time-varying point clouds used in AR applications \cite{NguyenCC:14,ThanouCF:16}. 

\section{System model}
\label{sec:SystemModel}
Let $V = \{V_1,\dots,V_N\}$ be the collection of aerial viewpoints captured by the UAVs. There are $K$ remote AR/VR sessions delivered to users interested in immersive visual navigation of different subareas (scenes) covered by the UAV fleet. Let $C_k$, for $k=1,\dots,K$, denote the subset of UAVs covering subarea $k$. For every UAV $\nu \in V$ (the association between captured viewpoints and UAVs is unique), let $R_\nu$ denote the temporal (data) sampling rate used in capturing viewpoint $\nu$. The aggregate transmission capacity that the UAV swarm can use to relay the captured data is $C$. The AR/VR application constructs a 3D 360$^\circ$-navigable video representation of the scene of interest for user $k$ from the subset of captured viewpoints $C_k$ for that subarea. Let $R^k = \{R_\nu | \nu \in C_k\}$ denote the set of sampling rates used by the UAVs from cluster $C_k$ to capture their respective viewpoints. Finally, let $\mathbf{R} = (R_1,\dots,R_N)$ denote the aggregate sampling rate vector.

We consider that each UAV will encode its captured viewpoint data $V_i$ into a scalable source-channel signal representation featuring $L$ embedded data layers and an aggregate data rate $R_i$. The description of this process and its optimization are described later in Section~\ref{sec:ScalableSourceChannel}. A UAV will then communicate its scalable signal representation towards the destination VR/AR user using its multi-beam directional antenna and optimal power scheduling/link assignment across the scalable signal representation layers. The description of this process is postponed for Section~\ref{sec:LayeredScheduling}.

\section{Space-time signal sampling/sensing}
\label{sec:ProblemFormulation}
Let $D_k$ denote the reconstruction error of scene $k$ observed by the remote VR/AR user. We define it as $D_k = \int_{v \in \mathcal{V}_k}w_v D_v(R^k)$, where $\mathcal{V}_k$ denotes the aggregate view space for the scene, $D_v$ is the reconstruction error for an arbitrary viewpoint $v \in \mathcal{V}_k$, and $w_v$ is the popularity (relevance) of viewpoint $v$ for the user\footnote{This quantity stems from the navigation pattern of the user for the scene.}. In words, $D_k$ represents the aggregate reconstruction error observed by the user across $\mathcal{V}_k$. Virtual (non-captured) views $v \in \mathcal{V}_k$ are synthesized for the user from the captured data $C_k$ via geometric signal processing\footnote{This operation requires scene depth signals that can be captured via IR sensors \cite{GokturkYB:04} or estimated from the captured viewpoints (color signals) \cite{TanimotoFS:07}.} (DIBR \cite{ShumCK:06}). Let $D_k^0$ represent an upper bound (constraint) on the reconstruction error for session $k$ imposed by the VR/AR application requirements.

The problem of interest can then be formulated as
\begin{align}
& \min_{\mathbf{R}}\sum_k \gamma_k D_k(R^k), \label{eqn:optimization_problem} \\
\text{subject to:} & \, \sum_i R_i \le C \, \text{and} \, D_k \le D_k^0, \forall k,
\nonumber
\end{align}

\noindent where $\gamma_k$ represents the prospective mission/application priority placed on session $k$ by the system. Note that $\mathbf{R}$ in essence samples the signal sensing locations $\{V_i\}$ across time and space, simultaneously. In particular, $R_i = 0$ signifies that signal/location $V_i$ is not sensed (at all). Note also that $D_k$ integrates the impact of the latency and interactivity requirements of the VR/AR application.

\section{Analysis and optimization}
\label{sec:AnalysisOptimization}
\vspace{-0.2cm}
In our recent work \cite{ChakareskiVS:12,VelisavljevicCC:11}, we have shown that when virtual view $v \in \mathcal{V}_k$ is interpolated via DIBR from the two nearest sampled viewpoints $V_i, V_{j} \in \mathcal{V}_k$, its reconstruction error $D_v(R^k)$ can be represented as a linear function of $D_{V_i}(R_i)$ and $D_{V_j}(R_j)$, each multiplied by polynomial powers of $x$, the relative position of $v$ with respect to them. 

Armed with this representation of $D_v(R^k)$, for virtual views $v$, we can reformulate the objective in \eqref{eqn:optimization_problem} by grouping terms associated with $D_i(R_i) = D_{V_i}(R_i), \forall i,$ to have

\begin{align}
& \min_{\mathbf{R}}\sum_i \alpha_i D_i(R_i), \label{eqn:optimization_problem2} \\
\text{subject to:} & \, \sum_i R_i \le C \, \text{and} \, D_k \le D_k^0, \forall k,
\nonumber
\end{align}

\noindent where the weights $\alpha_i \ge 0$ represent the aggregated terms.

Note that \eqref{eqn:optimization_problem2} now represents a convex optimization problem \cite{BoydV:03}. In particular, the weights $\alpha_i$ can be normalized to add to one, and the functions $D_i(R_i)$ can be accurately represented as concave functions $\exp(-\beta_i R_i)$ \cite{ChenPSLC:09}. Thus, \eqref{eqn:optimization_problem2} can be efficiently solved using convex optimization methods \cite{BoydV:03}.

\section{Scalable source-channel representation}
\label{sec:ScalableSourceChannel}
\begin{figure}[htb]
\centering
   \includegraphics[width=\figwidth]{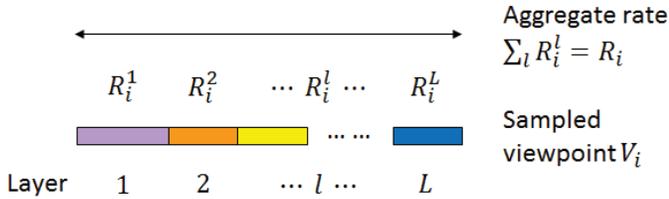}
   \vspace{-0.6cm}
   \caption{Scalable sampled viewpoint $V_i$ signal representation.}
   \label{fig:ScalableSourceCoding}
\end{figure}

To integrate the impact of unreliable/dynamic network links into the analysis, we design a scalable source-channel representation for the captured viewpoints $\{V_i\}$, as follows. First, the sampled $V_i$ data is encoded by the respective UAV into a set of embedded layers, illustrated in Figure~\ref{fig:ScalableSourceCoding}, featuring incrementally increasing levels of fidelity with the layer index $l$ \cite{SchwarzMW:07}. This will allow for inherent adaptation of the transmitted data to prospective bandwidth variations on the aerial network links, as the former is communicated towards the remote user.

\begin{figure}[htb]
\centering
   \includegraphics[width=\figwidth]{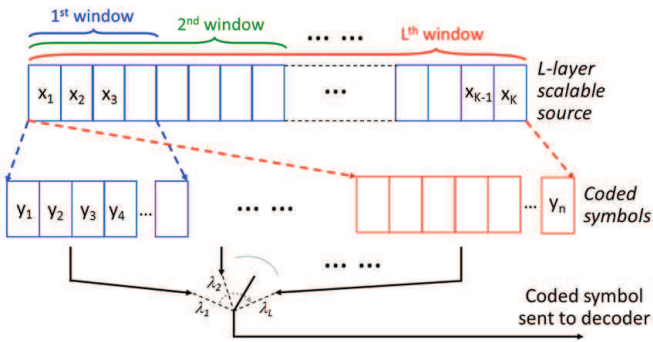}
   \vspace{-0.6cm}
   \caption{Rateless random linear channel coding.}
   \label{fig:RandomLinearCoding}
\end{figure}

Second, to protect against prospective transmission errors (packet loss) on the aerial links, each UAV applies efficient rateless random linear coding \cite{VukobratovicS:12} to its encoded viewpoint data, using the construction procedure from Figure~\ref{fig:RandomLinearCoding}.

In particular, $L$ transmission windows are constructed, where window $l$ represents the aggregate collection of the first $l$ layers of the encoded captured viewpoint. Coded symbols are constructed from each window using random linear combinations of the source symbols. Finally, a coded symbol is selected for transmission from window $l$ with probability $\lambda_l$.

Our embedded signal representation offers three advantages: inherent adaptability to dynamic network links, minimum required redundancy (to overcome transmission loss)\footnote{Our channel coding approach will allow UAV $i$ to maintain its aggregate source-channel data rate close to $R_i$, i.e., it will result in very little overhead.}, and inherent unequal error protection, all due to its design. The optimization problem that each UAV $i$ will consider here is
\begin{align}
& \min_{\{R_i^l\},\{\lambda_l\}}\sum_l E[D_i(\sum_{k=1}^l R_i^k),\{\lambda_l\}], \label{eqn:optimization_problem3} \\
\text{subject to:} & \, \sum_l R_i^l \le R_i \, \text{and} \, \sum_l \lambda_l = 1, \forall k,
\nonumber
\end{align}

\noindent where $R_i^l$ is the data rate of layer $l$ and $E[D_i(\sum_{k=1}^l R_i^k)|\{\lambda_l\}]$ is the expected reconstruction error of viewpoint $V_i$, when the first $l$ layers has been received/decoded by the destination VR/AR user/session\footnote{The probability of this event can be denoted as $P_l(\{\lambda_l\},R_i)$ \cite{VukobratovicS:12} and thus $E[D_i^l(\sum_{k=1}^l R_i^k)|\{\lambda_l\}] = P_l(\{\lambda_l\},R_i) D_i(\sum_{k=1}^l R_i^k)$.}.

Solving \eqref{eqn:optimization_problem3} is complex due to its nature. We design a coordinate descent algorithm to solve it iteratively, at low complexity. In particular, for fixed $\{\lambda_l\}$, \eqref{eqn:optimization_problem3} transforms into
\begin{align}
& \min_{\{R_i^l\}}\sum_l D_i(\sum_{k=1}^l R_i^k), \, \text{subject to:} \, \sum_l R_i^l \le R_i, \label{eqn:optimization_problem3_1}
\end{align}

\noindent which again is a convex optimization problem that can be solved efficiently. The algorithm operates by adjusting
$\{\lambda_l\}$ iteratively, in between every solution of \eqref{eqn:optimization_problem3_1}, until convergence. The adjustment is carried out such that the objective in \eqref{eqn:optimization_problem3_1} is minimized, every time $\{\lambda_l\}$ is updated. Formal description of our algorithm is provided in Algorithm~\ref{alg:local_search}.

\begin{algorithm}[htb]
\caption{Low-complexity coordinate descent}
\label{alg:local_search}
{\fontsize{9}{10}\selectfont
\begin{algorithmic}[1]
\STATE Initialize $\{\lambda_l\}=\{1,0,\dots,0\}$; Compute $\{R_i^l\}$ from \eqref{eqn:optimization_problem3_1}
\STATE Set $\Delta \lambda$; Assign $D_{max} = $ objective in \eqref{eqn:optimization_problem3}
\FOR{$i = 1$ to $L-1$}
\STATE FLAG1 = 0
\REPEAT
\STATE $\lambda_i=\lambda_i-\Delta\lambda; \lambda_{i+1}=\lambda_{i+1}+\Delta\lambda$
\STATE Assign $D = $ objective in \eqref{eqn:optimization_problem3}
\IF{$D < D_{max}$}
\STATE $D_{max} = D$
\STATE FLAG1 = 1
\ELSE
\STATE $\lambda_i=\lambda_i+\Delta\lambda; \lambda_{i+1}=\lambda_{i+1}-\Delta\lambda$
\STATE \textbf{break}
\ENDIF
\UNTIL{$\lambda_i \le \Delta\lambda$}
\IF{FLAG1 = 0}
\STATE \textbf{break}
\ENDIF
\ENDFOR
\end{algorithmic}
}
\end{algorithm}

Convergence of the algorithm is ensured, as the objective in \eqref{eqn:optimization_problem3} is positive, bounded from below by zero, and monotonically decreasing at every iteration.

\section{Layered directional networking}
\label{sec:LayeredScheduling}
We enhance the UAV transmission power/spectrum efficiency via smart multi-beam directional antennas \cite{AtmacaCE:09,CubicAntenna} and layered scheduling. Simultaneously, this will enhance the interactivity and quality of experience of the VR/AR applications, as it will reduce the transmission latency to them.

\begin{figure}[htb]
\centering
   \includegraphics[width=\figwidth]{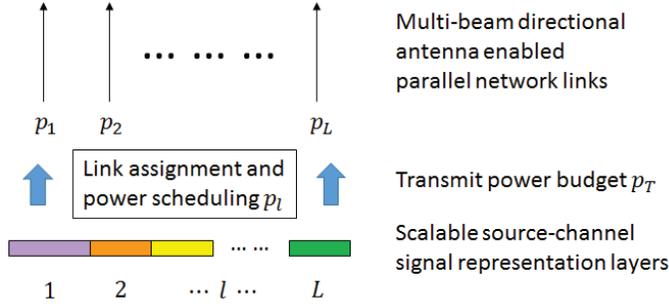}
   \vspace{-0.6cm}
   \caption{Optimal link assignment and transmit power scheduling.}
   \label{fig:LinkPowerScheduling}
\end{figure}

Let $L$ denote the number of beams/network links that the UAV antenna can establish towards its destination. Recall that $L$ also denotes the number of scalable signal representation layers constructed in Section~\ref{sec:ScalableSourceChannel}. Now, let $p_l$ denote the transmit power allocation to link $l$. Coded symbols from transmission window $l$ in Section~\ref{sec:ScalableSourceChannel} are assigned to the corresponding network link for transmission. UAV $i$ assigns its power budget $p_T$ according to the following optimization
\begin{align}
& \min_{\{p_l\}}\sum_l p_l, \, \text{subject to:} \, \sum_l p_l \le p_T \, \text{and} \, r(p_l) = R_i^l, \forall l, \label{eqn:optimization_problem4}
\end{align}

\noindent where $r(p_l)$ denotes the data rate on link $l$ enabled by $p_l$. \eqref{eqn:optimization_problem4} also represents a convex optimization problem that can be solved efficiently. The optimization is illustrated in Figure~\ref{fig:LinkPowerScheduling}.

Note that we consider that $p_T$ is much larger than the minimum required power to transmit the aggregate sampling rate $R_i$ of UAV $i$. This is a reasonable assumption that any UAV battery can easily meet, given its much higher energy volume. In particular, the energy consumption for UAV propulsion is a few orders of magnitude larger than that for sensing/communication.

\section{Experiments}
\label{sec:Experiments}
We carry out experiments to evaluate the effectiveness of our framework. They involve aerial viewpoint data captured in both, a controlled laboratory setting and an outdoor (in-the-wild) environment \cite{VIRAT}. For reproducibility and scientific analysis, our experimentation methodology was carried as follows. Viewpoint data was acquired using DJI Phantom 4 UAVs equipped with 4K cameras and 4G LTE dongles \cite{DJI-UAV}. Since this is a proof-of-concept study of a novel application and given the limited space here, we only considered a one-hop aerial network in our experiments. That is, each UAV is directly connected to a ground-based base station to which it sends its data. We collected measurements of the wireless link to the base station for different UAV locations. These traces and the UAV captured data were then used to apply our optimization framework components via numerical simulations. We set $K = 6$ equal priority ($\gamma_k = 1$) VR/AR sessions delivered remotely. Their subareas/scenes of interests are spatially distributed across the area under the UAV swarm such that every $C_k$ comprises at least three captured viewpoints $V_i$ covering it. The viewpoint popularity distribution $w_v$ for scene $C_k$ was selected to be Gaussian over $\mathcal{V}_k$, matching the field of view of the VR/AR head-mounted displays.

We designed a reference system to compare to, as follows. The UAV swarm samples the captured viewpoints at equal rate $R_i = C/N$ (uniform allocation). Conventional state-of-the-art video encoding is employed to represent the captured data at each UAV \cite{WiegandSBL:03}. Conventional state-of-the-art (Reed-Solomon) forward error correction is applied to the captured data at fixed rate (3\%) \cite{LinC:84}. The reference system is denoted as {\em Baseline} and our optimization framework as {\em Optimal}, henceforth.

\begin{figure}[htb]
\centering
   \includegraphics[width=\figwidth]{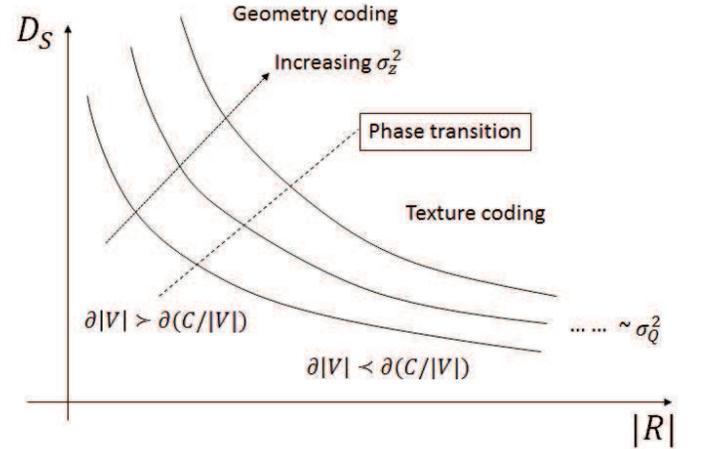}
   \vspace{-0.8cm}
   \caption{Scaling behavior of the optimal sampling vector $R$.}
   \label{fig:ScalingBehavior}
\end{figure}

{\bf Scaling behavior.} In Figure~\ref{fig:ScalingBehavior}, we study how the optimization \eqref{eqn:optimization_problem} behaves as we scale $C$. In particular, let $D_S$ denote the objective function in \eqref{eqn:optimization_problem}. We record its values for the optimal sampling rate vector $R$, as we vary $C$. We plot the resulting dependency $D_S$ versus $|R|$ (where $|R| \sim C$), parameterized by $\sigma_z$, the standard deviation of the surface $f(x,y)$ representing the 3D geometry of the area under the UAV swarm (we can control this quantity in laboratory settings).
We can observe two distinct operating regimes (modes). For smaller $|R|$, $D_S$ scales as $c_1(\sigma_z^2)|R|^{-c_2(1/\sigma_z^2)}$, where $c_1$ and $c_2$ are two constants that depend on $\sigma_z$. We denote this mode {\em geometry coding}, since the optimal sensing policy emphasizes the fidelity of the representation of $f(x,y)$ here, due to its impact on $D_S$. This is accomplished by prioritizing sampling more viewpoints $\nu \in V$ at the expense of the data/temporal sampling rate assigned to each such $\nu$. Note that the higher the innovation rate (or entropy) of $f(x,y)$ across the imaged area, as embodied by $\sigma_z$, the higher $D_S$ is, as seen from Figure~\ref{fig:ScalingBehavior}. Once $|R|$ is sufficiently large and $f(x,y)$ can be represented at sufficiently high fidelity, we observe a transition to another operating regime, where $D_S$ now scales as $c_3(\sigma_c^2)|R|^{-c_4(1/\sigma_c^2)}$, and $c_3$/$c_4$ are again constants that depend on $\sigma_c$ (the standard deviation of the scene color values). We denote this mode {\em texture coding}, as the optimization emphasizes here the color value representation fidelity for every triple $(x,y,f(x,y))$. This is accomplished by emphasizing temporal sampling at the expense of the number of sampled viewpoints. Still, the impact of $\sigma_c$ is much smaller here and all three $D_S$ versus $|R|$ graphs eventually converge\footnote{$\sigma_Q$ is the quantization step size used to encode the captured $\{V_i\}$ \cite{GershoG:91}.}.

\begin{figure}[htb]
\centering
   \includegraphics[width=\figwidth]{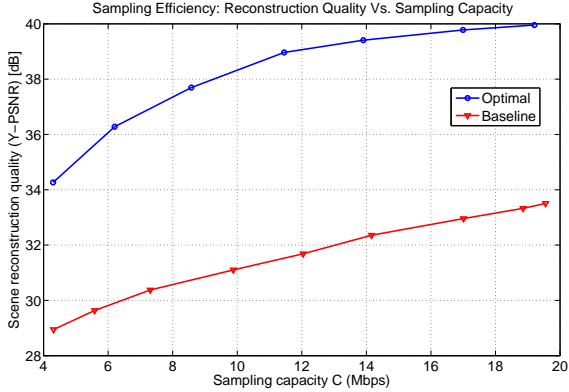}
   \vspace{-0.8cm}
   \caption{Reconstruction quality versus sampling capacity (outdoors).}
   \label{fig:SamplingEfficiencyOutdoor}
\end{figure}
{\bf Sampling efficiency.} In Figure~\ref{fig:SamplingEfficiencyOutdoor}, we examine the sampling efficiency of our framework relative to the baseline reference. In particular, we graph the achieved expected scene reconstruction quality (the inverse of the objective in \eqref{eqn:optimization_problem}) versus the available sensing capacity, for the two methods. It can be seen that the optimization is able to leverage the available resources much more effectively, enabling considerable reconstruction quality gains over the baseline system, for the same $C$. Furthermore, we can see that {\em Optimal} is able to upscale its performance at much higher rate as $C$ is increased, relative to {\em Baseline}. The equivalent results for the indoor data look similar. To conserve space, we do not include them here.

\begin{figure}[htb]
\centering
   \includegraphics[width=\figwidth]{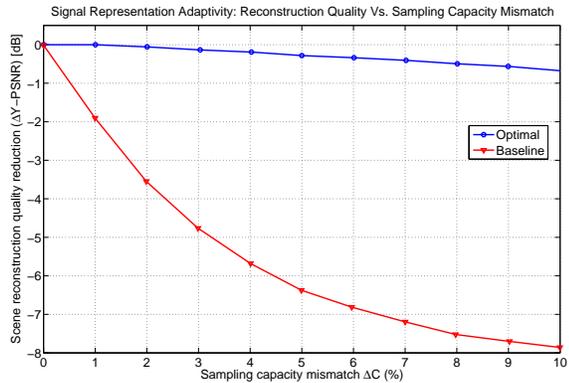}
   \vspace{-0.8cm}
   \caption{Signal representation adaptivity versus capacity mismatch $\Delta C$ (\%).}
   \label{fig:SignalRepresentationAdaptivity}
\end{figure}
{\bf Signal representation adaptivity/reliability.} We carry out two experiments here to examine the adaptivity of our signal representation design to sampling capacity mismatch and its robustness to unreliable transmission, respectively. In particular, in the first experiment, we consider that the captured viewpoint data $\{V_i\}$ has been encoded for target capacity $C_0$ (12 MBps here), however, the actual available capacity $C$ at which it can be relayed to the destination VR/AR sessions is smaller. We measure the expected scene reconstruction quality achieved thereby, for different values of the mismatch $\Delta C$, expressed in percent of $C_0$. These results are shown in Figure~\ref{fig:SignalRepresentationAdaptivity}. We can see the our approach offers a graceful degradation as $\Delta C$ is increased. In fact, our signal design does not exhibit any notable reduction in representation efficiency had the captured data been encoded instead for target sampling capacity $C - \Delta C$, as the comparison with the corresponding graph from Figure~\ref{fig:SamplingEfficiencyOutdoor} shows. On the other hand, even a small mismatch can degrade the end-of-the-end performance of the baseline method considerably, as is evident from Figure~\ref{fig:SignalRepresentationAdaptivity}, due to the lack of adaptability of the conventional signal representation it employs.

\begin{figure}[htb]
\centering
   \includegraphics[width=\figwidth]{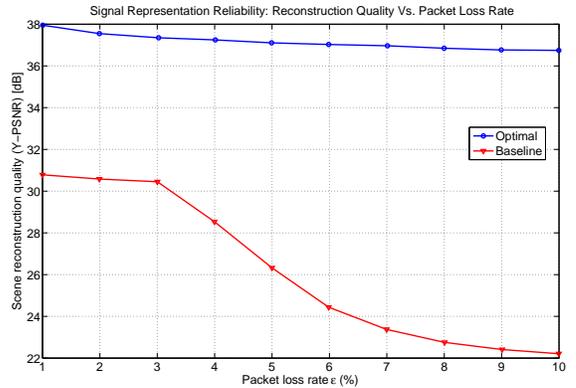}
   \vspace{-0.8cm}
   \caption{Signal representation reliability versus packet loss rate $\epsilon$ (\%).}
   \label{fig:SignalRepresentationReliability}
\end{figure}
The second experiment measures the achieved reconstruction quality across the VR/AR sessions as the average packet loss rate $\epsilon$ on the aerial network links is increased. These results are shown in Figure~\ref{fig:SignalRepresentationReliability}. We can see that our source-channel signal representation offers robustness and intrinsic adaptation to varying packet loss, without penalizing the end-to-performance. On the other hand, the baseline reference exhibits a cliff-effect, typical for conventional error protection \cite{RaneBG:08}, beyond which the reconstruction quality dramatically degrades, as the packet loss increases, as shown in Figure~\ref{fig:SignalRepresentationReliability}.

{\bf Transmission power/latency reduction.} We measured in our experiments that our framework consumes approximately 30\% less transmission power relative to the baseline reference. This observation is in line with results reported in prior studies of directional networking \cite{SpyropoulosR:02}. Simultaneously, we also observed that the delivery latency for the VR/AR applications has been reduced by 25\%. These advances make the practical deployment of our framework more feasible.
\vspace{-0.01cm}

\section{Conclusion}
\label{sec:Conclusion}
\vspace{-0.05cm}
We explored UAV IoT aerial sensing for VR/AR immersive communication to remote users in a first-of-its kind study on a novel topic of prospectively broad societal connotations. The remote users are interested in visual immersive navigation of specific subareas/scenes of interest, reconstructed on their respective VR/AR devices from the aerial viewpoint data captured by the UAV swarm. We designed an optimization framework that comprises three components that synergistically operate to maximize the end-to-end performance across the VR/AR sessions delivered to the users. In particular, we first derived the optimal sampling rates to be used by each UAV, such that the related system and application constraints are not exceed, while the priority weighted reconstruction quality across all VR/AR sessions is maximized. Then, we formulated an optimal scalable source-channel signal representation that imbues into the captured data inherent rate adaptivity, unequal error protection, and minimum required redundancy. Finally, we enhanced the UAV transmission efficiency via small-form-factor multi-beam directional antennas and optimal power/link scheduling across the scalable signal representation layers. Our experiments demonstrate competitive advantages over conventional methods employed in practice for visual sensing. The demonstrated advances are very promising and motivate us to proceed with a follow-up study, in which we plan to pursue closer integration of our optimization methods and further analysis of their operation.

\bibliographystyle{IEEEtran}
\bibliography{c://Jakov/Publications/_StyleFiles/myrefs}

\end{document}